\documentclass[twocolumn, superscriptaddress,showpacs,preprintnumbers,amsmath,amssymb]{revtex4}
\usepackage{graphicx}
\usepackage{bm}

\newcommand{\vect}[1]{\bm{#1}}
\newcommand{\deffig}[4]{
 \begin{figure}
 \includegraphics[scale=#3]{#2}
 \caption{\label{fg:#1} #4}
 \end{figure}
}
\newcommand{\beq}{\begin{equation}}
\newcommand{\eeq}{\end{equation}}
\newcommand{\Fig}[1]{Fig.\,\ref{fg:#1}}

\newcommand{\Eq}[1]{Eq.(\ref{eq:#1})}

\newcommand{\Prob}{{\rm Prob}}

\begin{document}


\title{
  Ground States of the SU($N$) Heisenberg Model
}
\author{
  Naoki~Kawashima and Yuta~Tanabe
}
\affiliation{
  Institute for Solid State Physics, University of Tokyo,
  Kashiwa 5-1-5, Chiba 277-8581, Japan
}

\date{\today}


\begin{abstract}
The SU($N$) Heisenberg model with various single-row representations 
is investigated by quantum Monte Carlo simulations.
While the zero-temperature phase boundary agrees qualitatively 
with the theoretical predictions based on the $1/N$ expansion, 
some unexpected features are also observed.
For $N \ge 5$ with the fundamental representation, for example,
it is suggested that
the ground states possess exact or approximate U(1) degeneracy.
In addition, for the representation of Young tableau with 
more than one column, the ground state shows no VBS order
even at $N$ greater than the threshold value.
\end{abstract}


\maketitle


Since the resonating-valence-bond (RVB) state
was proposed\cite{Anderson} as a possible mechanism 
that supports novel super conductivity in cupurates,
it has been a major target of condensed matter theory
to find a short-range interaction model that realizes 
a spin liquid state at zero temperature.
Introducing frustrations into the model that otherwise
has a magnetic ground state is one of promising directions
to achieve this goal.
Numerical verification for such models is, 
however, technically very difficult,
and conclusive results are still missing
for geometrically frustrated systems such as the 
antiferromagnet on a triangular lattice.
Another approach was taken in \cite{ReadS1989,ReadS1990}, 
where the authors generalized the Heisenberg 
antiferromagnet to higher symmetries, 
thereby increasing the model's degrees of freedom 
and enhancing quantum fluctuations.
Based on the $1/N$ expansion treatment, they predicted
that the model with sufficiently large $N$ has a valence-bond-solid
(VBS) ground state with spontaneously-broken lattice symmetry.
The nature of the ground states may depend on the
representation, somewhat analogous to what happens in the 
SU(2) models in one dimension 
though the underlying mechanisms may be rather different.

\deffig{VBS}{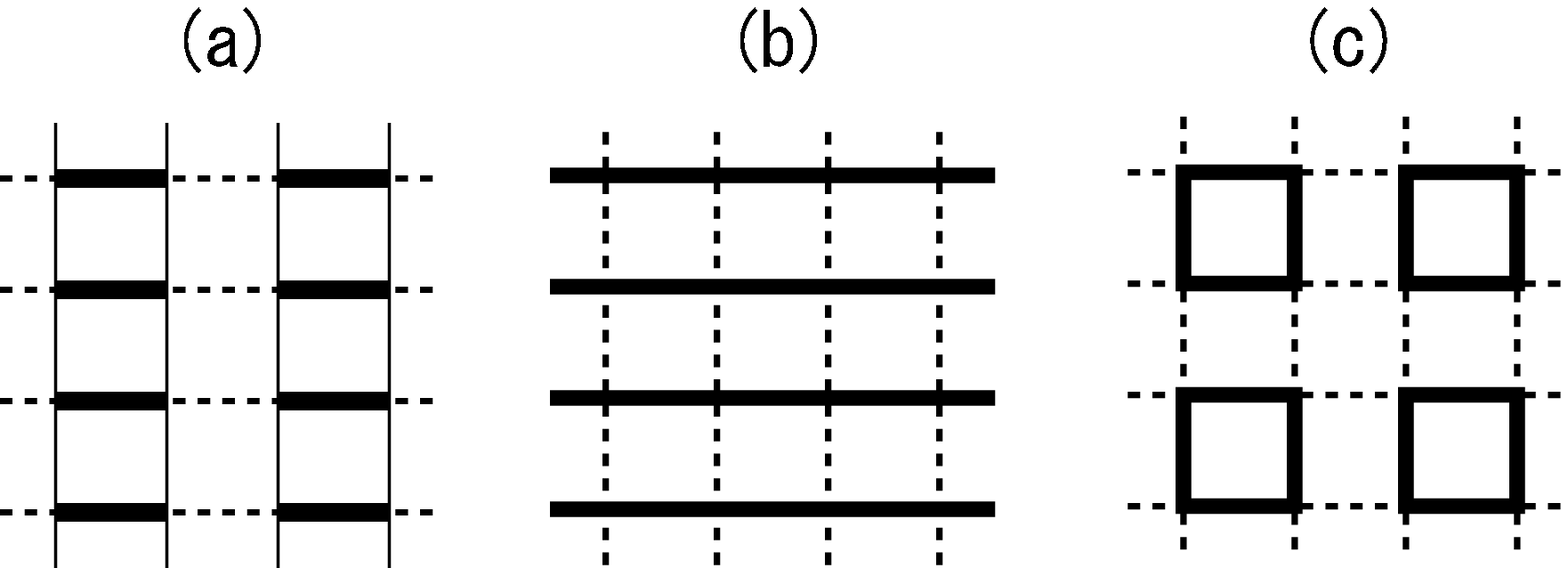}{0.45}
{Three types of VBS states: 
(a) columnar ($\overline{D_x} \ne 0$ and $\overline{D_y} =   0$),
(b) nematic ($\overline{D_x} = \overline{D_y} = 0$),
(c) plaquette ($\overline{D_x} = \overline{D_y} \ne 0$).
The quantities $\overline{D_x}$ and $\overline{D_y}$ characterize
spontaneous translational-symmetry-breaking in $x$ and $y$ 
directions, respectively.
(See \Eq{DefinitionOfD} for the definition.)
}

More specifically, it was suggested\cite{ReadS1989,ReadS1990} that,
for the model with the Young tableaux with $m$ rows and $n$ columns,
the $N$-$n$ phase diagram does not strongly depend on $m$, and
has a line of phase transition separating the small-$N$ 
N\'eel region from the large-$N$ VBS region.
It was also argued that the nature of the
VBS ground state can be classified according to 
the quotient of the division of $n$ by 4.
If $n \equiv 1$ or $3$ (mod 4), the ground state has a columnar 
ordering (\Fig{VBS} (a)) with the translational symmetry and
the 90-degree rotational symmetry both broken,
whereas if $n \equiv 2$ (mod 4) it has a ``nematic'' VBS 
ordering (\Fig{VBS} (b)) with only the lattice-rotational 
symmetry broken.
Finally, if $n$ is a multiple of 4,
there is no spontaneous breaking of the lattice symmetry.

They also showed that the $O(1/N)$ effective 
Hamiltonian with the fundamental representation is exactly 
the same as the quantum dimer model 
\cite{RokhsarK1988} at $V=0$.
For the quantum dimer model, it was shown
by a numerical calculation\cite{RKnumerical} 
that the ground state is a VBS state with columnar arrangement of dimers.
This is in agreement with the above conjecture that 
the degeneracy is 4 for $n=1$.
A direct check of the spontaneous breaking-down of 
the translational symmetry for $m=n=1$ was
carried out in the previous paper \cite{HaradaKT2002}, 
which yielded the transition value of $N$, 
namely $4<N^{\ast}(m=n=1)<5$.

The model Hamiltonian we discuss is defined as
\begin{equation}
  H = J 
  \sum_{(\vect{R},\vect{R}')}
  \sum_{\alpha,\beta=1}^{N} 
  S^{\alpha\beta}(\vect{R}) S^{\beta\alpha}(\vect{R}')
  \label{eq:Hamiltonian}
\end{equation}
where the operator $S^{\alpha\beta}$ is the generator
of the SU($N$) algebra.
Here we consider the simple square lattice with the
periodic boundary condition.
We divide the whole lattice into to two sublattices, say, A and B.
The representation of the generators on sublattice A is 
characterized by the Young tableau with a single raw ($m=1$) 
and varying number of columns.
The representation on sublattice B is the conjugate of that 
on sublattice A.
We have performed quantum Monte Carlo simulations by 
the directed loop algorithm\cite{SyljuasenS2002} and 
extrapolated the results to the zero temperature limit.
The algorithm can be obtained by generalizing the idea 
of the coarse-grained algorithm for $N=2$ \cite{HaradaK2002} 
to the present general $N$ case.
The technical details will be published elsewhere.\cite{Future}

Our principal findings are as follows:
(i) For $n=1$, the ground state in the large-$N$ region adjacent 
to the N\'eel region is the columnar VBS state 
as the previous works suggested.
However, up to $L=32$, it appears to have
infinite degeneracy corresponding to a continuous U(1)
symmetry that does not exist 
in the original microscopic Hamiltonian 
but may emerge asymptotically.
Accordingly, the resonons \cite{RokhsarK1988} are gapless.
(ii) The ground state in the large-$N$ region
for $n>1$ shows no evidence for break down of any 
lattice translational symmetry, in particular, 
the ones predicted in Read and Sachdev\cite{ReadS1989,ReadS1990}.
The distance dependence of the energy-energy correlation function 
is consistent with the algebraic decay.


\deffig{M1RMX}{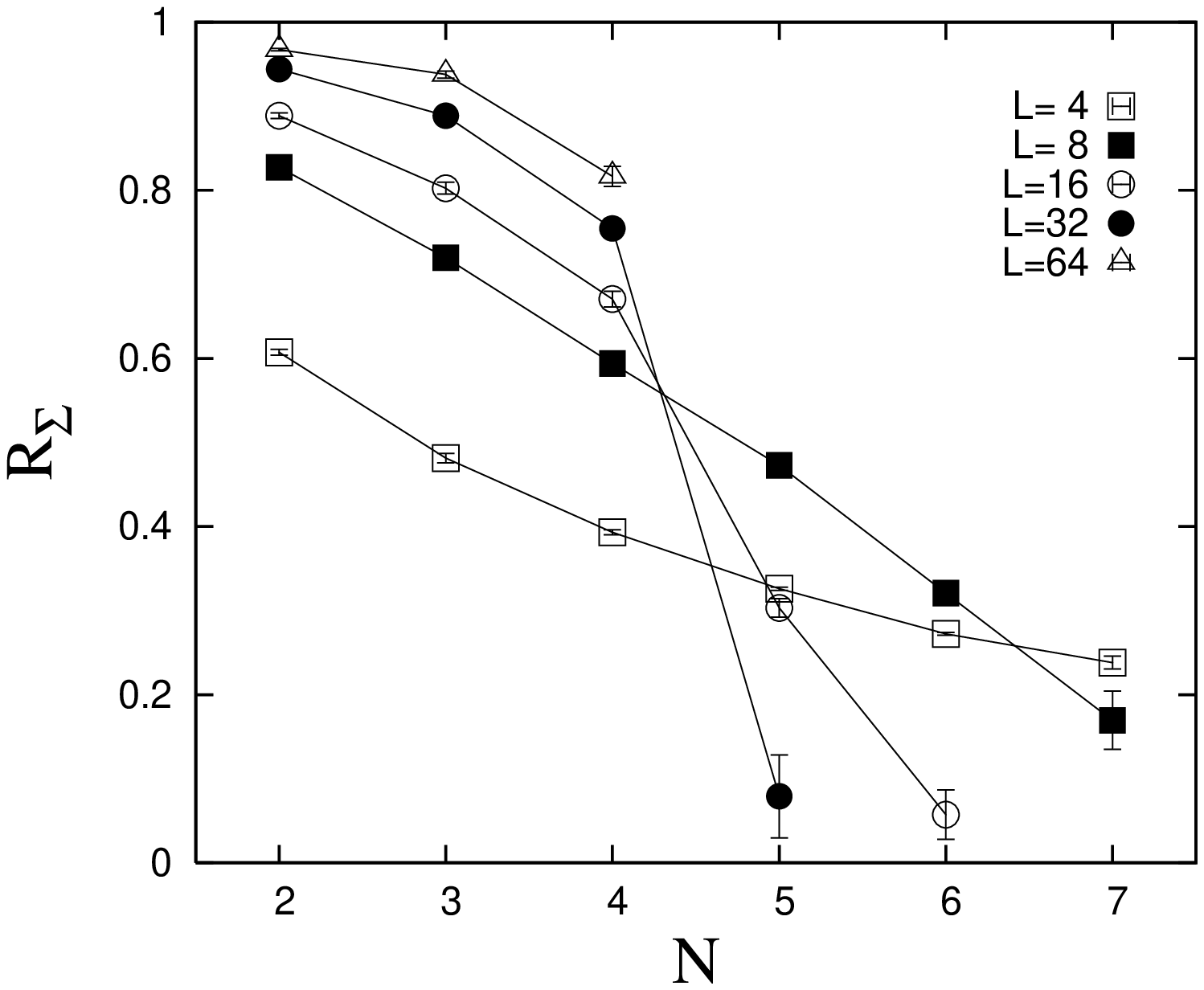}{0.40}
{The correlation ratio, $R_{\Sigma}(L)$, of the magnetization for 
the fundamental representation ($n=1$).}

In the present letter, we start from showing some
results that confirm the conclusions of
the previous numerical work on the SU($N$) model,
and then discuss the new findings.
We first look at the case of the fundamental representation ($n=1$).
It was found\cite{HaradaKT2002} that 
for $N\le 4$ the ground state is a N\'eel state
whereas it does not have a spontaneous staggered magnetization for $N \ge 5$. 
Instead the ground state for $N=5$ and $6$
was found to possess the spontaneous dimerization.
Although a macroscopic quantity is often used for
a probe for detecting the transition,
here we examine two-point correlation functions
in order to see the transition more clearly,
$
  C_Q(R;L) \equiv 
  \langle Q(R\vect{e}_{\mu}) Q(\vect{0}) \rangle_L 
  - \langle Q(R\vect{e}_{\mu}) \rangle \langle Q(\vect{0}) \rangle_L
$
where $L$ is the system size, 
$\vect{e}_{\mu}$ \, $(\mu=x,y)$ a lattice unit vector,
and $Q(\vect{R})$ an arbitrary quantity locally defined 
around the position $\vect{R}$.
We also use the correlation ratio\cite{TomitaO2002}
defined simply as 
$
  R_Q(L) = C_Q(L/2;L)/C_Q(L/4;L).
$
Similar to the Binder parameter,
the correlation ratio is a dimensionless quantity and, 
when plotted against a relevant physical parameter,
the common crossing point of curves with various system sizes
marks the transition point.
In \Fig{M1RMX}, the correlation ratio for the ``magnetic'' moment
is plotted against $N$ for various system sizes for $m=1$.
Here, the ``magnetic'' moment is defined as 
$$
  \Sigma(\vect{R}) 
  \equiv S^{11}(\vect{R}) - S^{22}(\vect{R}).
$$
While the ratio tends to converge to
unity for $N=2,3$ and $4$, indicating the existence of the
N\'eel ordering, it rapidly decreases for $N\ge5$.
Based on this figure, we can estimate the transition point as
$N^{\ast}(n=1) \sim 4.3$.


While \Fig{M1RMX} establishes the absence of the 
magnetic ordering for $N=5,6,7$, it does not tell us 
much about the nature of the ground state.
It was shown in \cite{HaradaKT2002} that in the
ground state the lattice translational invariance is broken.
Here we reconfirm the presence of the VBS order
by detecting the long-range correlation in $C_A(\vect{R};L)$ 
where $A$ is defined as
\begin{equation}
  A(\vect{R}) 
  \equiv 
  P_x(\vect{R}) - P_y(\vect{R}),
\end{equation}
and $P_{\mu}(\vect{R})$ is the nearest-neighbor product 
of the ``magnetic'' moments,
\begin{equation}
  P_{\mu}(\vect{R})
  \equiv
  \Sigma(\vect{R}) \Sigma(\vect{R}+\vect{e}_{\mu}).
\end{equation}
In \Fig{M1RST}, $C_A(L/2;L)$ is plotted.
It is expected that the correlation must decay down to zero 
in the large $L$ limit for a N\'eel state,
whereas for any type of the VBS states in \Fig{VBS} it must 
converge to a non-zero value.
\Fig{M1RST} clearly shows that the large-$N$ phase for the
fundamental representation is the VBS phase.

\deffig{M1RST}{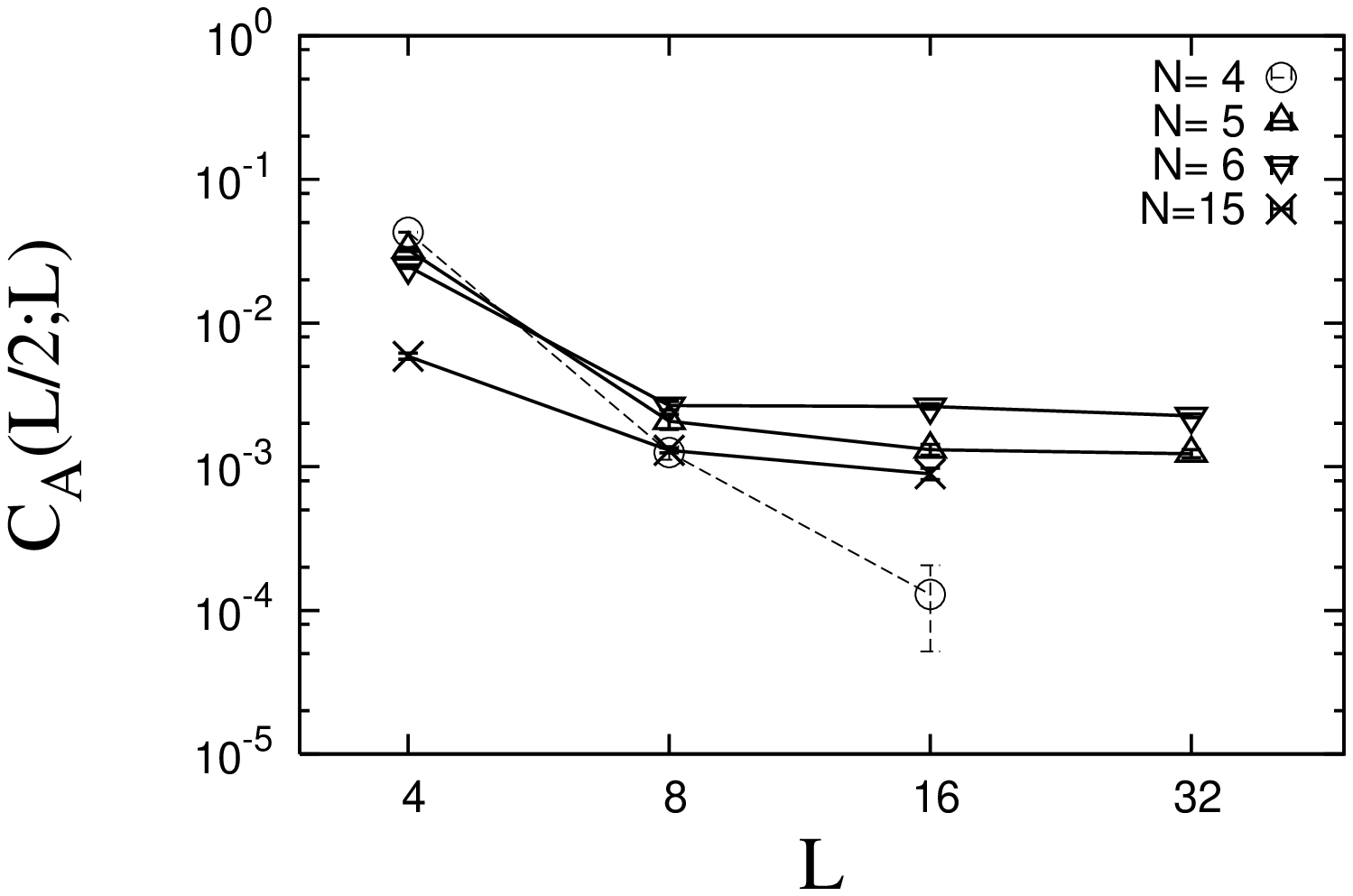}{0.5}
{
The two-point correlation function the rotational-symmetry breaking
order parameter, $A$, for the model with the fundamental 
representation ($n=1$).
}


These observations are consistent with 
the columnar VBS states (such as (a) in \Fig{VBS}) 
as is conjectured for $N>N^{\ast}$ \cite{ReadS1989,ReadS1990}.
However, the above-mentioned probe $A$ only is not sufficient to
rule out the plaquette VBS state ((c) in \Fig{VBS}).
Since the columnar VBS states do not possess the 90-degree 
rotational symmetry while the plaquette VBS state does, 
one may naively expect that measuring the difference between
the average bond strength in the $x$ direction and that 
in the $y$ direction can distinguish the two.
Therefore, we have computed the quantity
$
  B \equiv 
  \langle \left( \overline{P_x} - \overline{P_y} \right)^2 \rangle
$
where an overline indicates the average over the volume, i.e.,
$
  \overline{P_{\mu}} \equiv V^{-1} \sum_{\vect{R}} P_{\mu}(\vect{R}).
$
The quantity $B$ is plotted in \Fig{AverageBondDifference} against the
system size.

\deffig{AverageBondDifference}{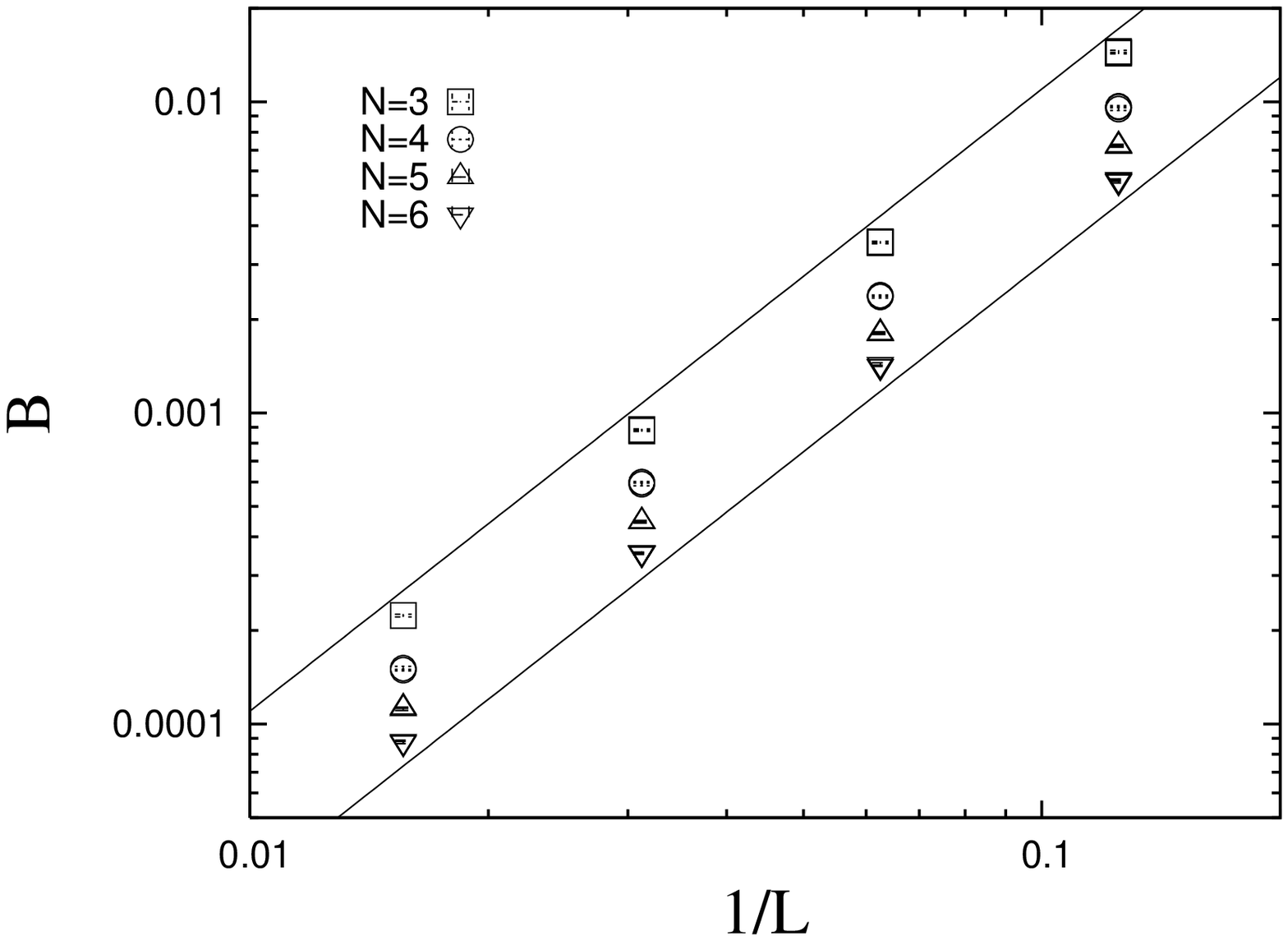}{0.40}
{
  The squared difference of the average bond strengths in the
  $x$ and $y$ direction for the fundamental representation.
  The straight lines that correspond to $B \propto 1/L^2$
  are drawn for comparison.
}

Instead of distinguishing the two types of states,
the quantity $B$ turns out to reveal an interesting property.
The quantity $B$ is proportional to $1/L^2$ not only 
for $N=3$ and $4$ but also for $N=5$ and $6$, 
indicating that the average bond strength in the $x$ direction 
is the same as that in the $y$ direction even in the VBS state.
It follows that the expectation value of the nearest-neighbor correlation
$P_{\mu}(\vect{R})$ in a single (pure) VBS ground state has the form
\begin{equation}
  \langle P_{\mu}(\vect{R}) \rangle_{\rm single}
   = \overline{P} + \overline{D_{\mu}}\times (-1)^{R_{\mu}}
  \quad (\mu = x,y)
  \label{eq:DefinitionOfD}
\end{equation}
with the direction-independent average value $\overline{P}$ 
and some direction-dependent constant
$\overline{D_{\mu}}$ that characterizes the dimerization order.
The constants $\overline{D_x}$ and $\overline{D_y}$ 
are expressed by the local quantities $D_{\mu}(\vect{R})$ as
$
  \overline{D_{\mu}} = V^{-1} \sum_{\vect{R}} D_{\mu}(\vect{R})
$
where
$
  D_{\mu}(\vect{R}) 
  \equiv
  \left( 
  P_{\mu}(\vect{R+\vect{e}_{\mu}}) - P_{\mu}(\vect{R})
  \right) / 2.
$

In order to distinguish the columnar state from the plaquette state,
we have to examine the joint distribution function,
$\Prob(\overline{D_{x}},\overline{D_{y}})$.
For small systems, it consists of a single broad peak at the center,
$(\overline{D_x},\overline{D_y}) = (0,0)$, even for $N \ge 5$.
However, as the system grows the weight at center should diminish and
peaks must appear according to the structure of the VBS phase.
If the columnar VBS states are the true ground state, 
four peaks must develop at symmetric positions on $x$ and $y$ axes 
( $(\overline{D_x},\overline{D_y}) = (\pm D,0), (0,\pm D)$ )
whereas if the plaquette VBS states are the ground states
the peaks must appear at diagonal positions 
( $(\overline{D_x},\overline{D_y}) = \pm (D,D), \pm(D,-D)$ ).

\deffig{Circle}{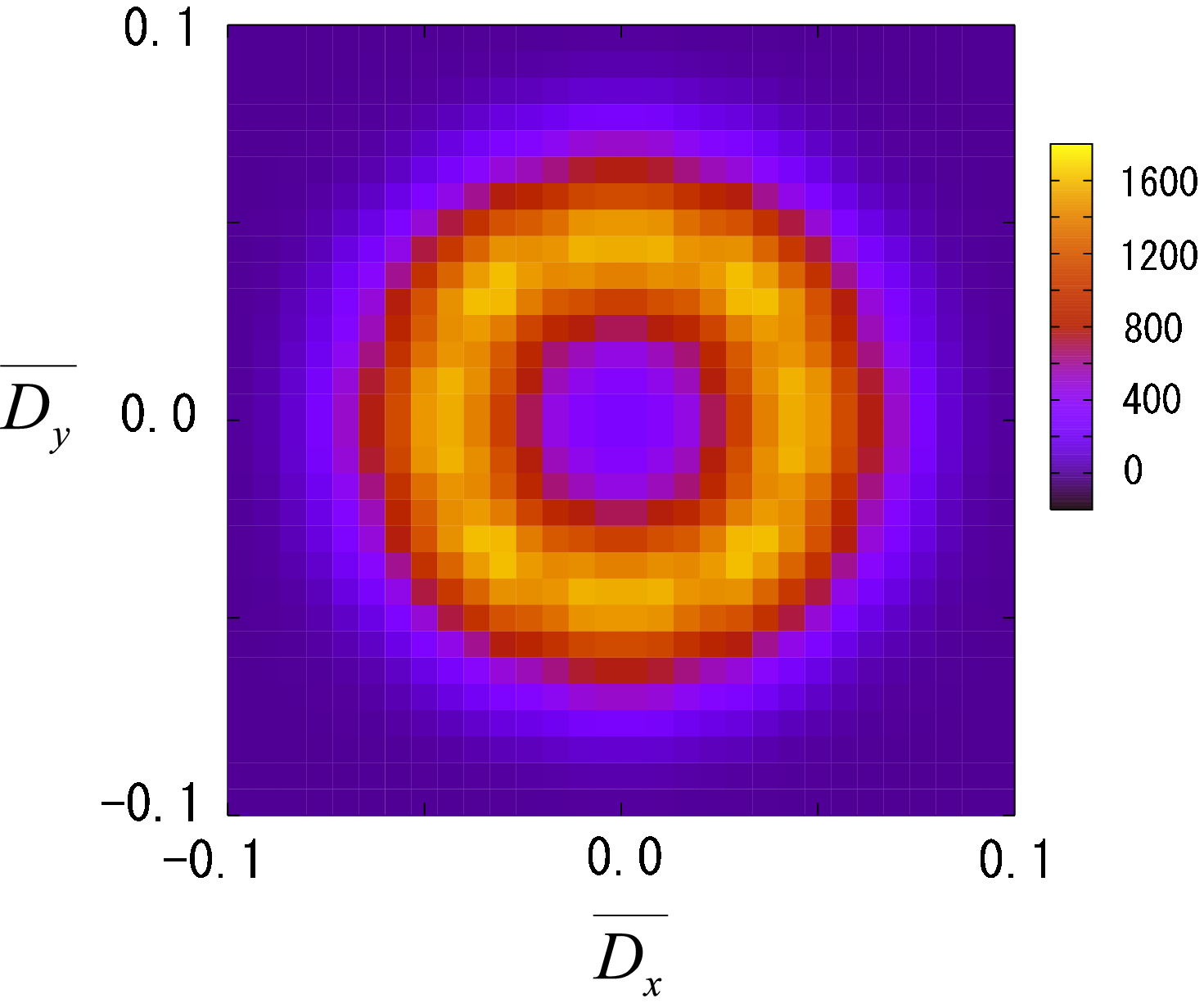}{0.45}
{
The frequency of observing a pair of values 
$(\overline{D_x},\overline{D_y})$ 
during the Monte-Carlo simulation for $N=6$ at $L=32$ and $\beta=16$.
The value at each pixel is the average over eight 
symmetrically equivalent points, 
$(\pm\overline{D_x}, \pm\overline{D_y})$ and
$(\pm\overline{D_y}, \pm\overline{D_x})$ 
}

As we see in \Fig{Circle},
the distribution obtained from our computation shows neither feature.
The distribution is circularly symmetric.
This feature does not depend on $N$, the temperature, or the system size
(at least up to $L=32$), whenever a finite VBS order is observed.
This suggests that the ground state is not 
only four-fold degenerate but infinitely degenerate.
The ground state manifold possesses
the U(1) invariance although it is not obvious from 
the microscopic Hamiltonian \Eq{Hamiltonian}.
Since each ground state breaks this U(1) symmetry, there must be a 
Goldstone mode associated with it, which may correspond to
``resonons'' mentioned in \cite{RokhsarK1988},
though its gapless nature was predicted only for an isolated 
critical point.
If the apparent U(1) symmetry persists to the thermodynamic limit,
the resonons must be gapless not only at the isolated critical point
$N=N^{\ast}$ but also in the whole region of the VBS phase.
However, another possibility cannot be excluded by the present numerical 
calculation, i.e., the possibility that this apparent U(1) symmetry may be
a transient behavior that applies only to certain intermediate length 
scales and the discrete symmetry of the original microscopic 
Hamiltonian is recovered for larger length scales.
If this is the case, there should be a finite number of
preferred angles in the two-component order-parameter space
to which the true ground states correspond,
and the resonon modes have a small but finite excitation gap.



\deffig{M2RMX}{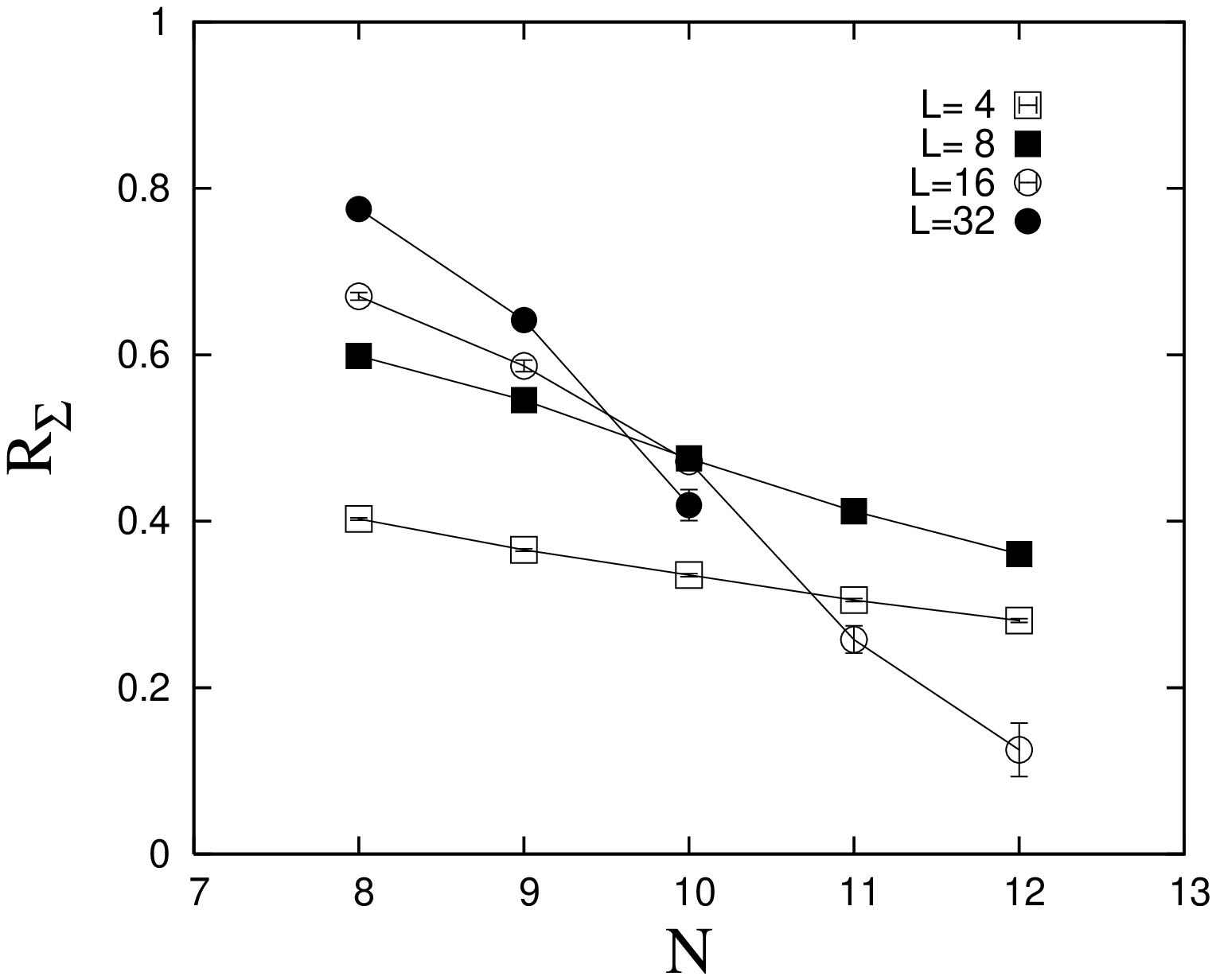}{0.40}
{
The correlation ratio of the magnetization for the representation of
two-box Young tableau ($n=2$).
}

Next we consider the model with the representation of Young's tableaux
with two or more columns ($n \ge 2$).
In order to determine the critical value of $N$, we have computed the
correlation ratio for the magnetization as described above.
For $n=2$, the correlation ratio for the magnetization is
plotted in \Fig{M2RMX}.
The N\'eel state is the ground state for $N \le 9$
whereas it is not for $N=10$ and larger.
We can see the trend changes around $N^{\ast}(n=2) \sim 9.5$ at which
curves cross each other.
This is again in a good agreement with $1/N$ expansion result, 
$N^{\ast} \sim 5.3\, n$ \cite{ReadS1990,ArovasA1988}.
For $n=3$ and $n=4$ we have performed similar computation and located
the the phase boundary as $N^{\ast}(n=3) \sim 14$ and $N^{\ast}(n=4) \sim 20$,
respectively.
These results are summarized in the schematic phase diagram (\Fig{PhaseDiagram}).

\deffig{PhaseDiagram}{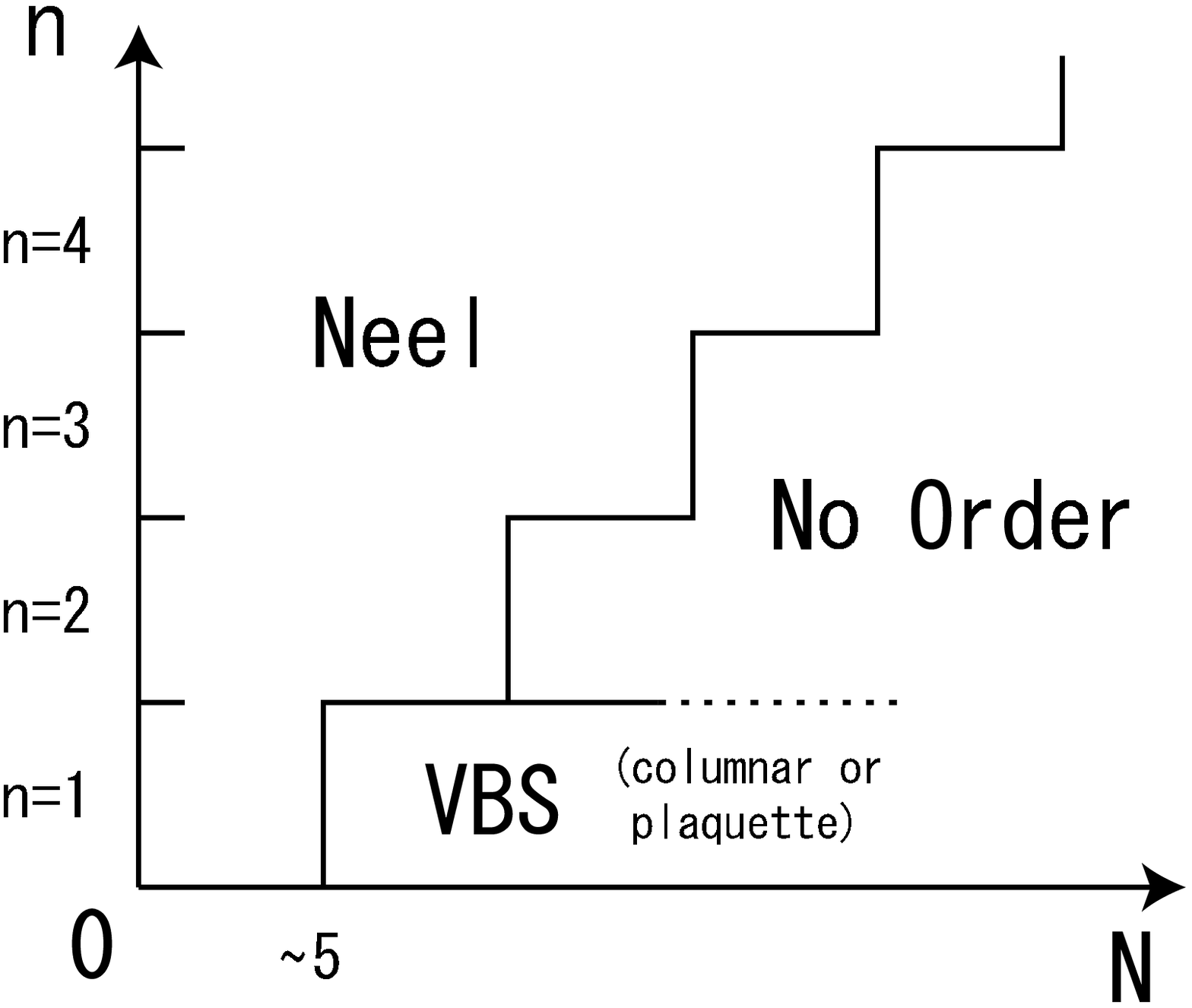}{0.35}
{The zero-temperature phase diagram of SU($N$) model on a square lattice
with single-row ($m=1$) representations.}


\deffig{CST_vs_L}{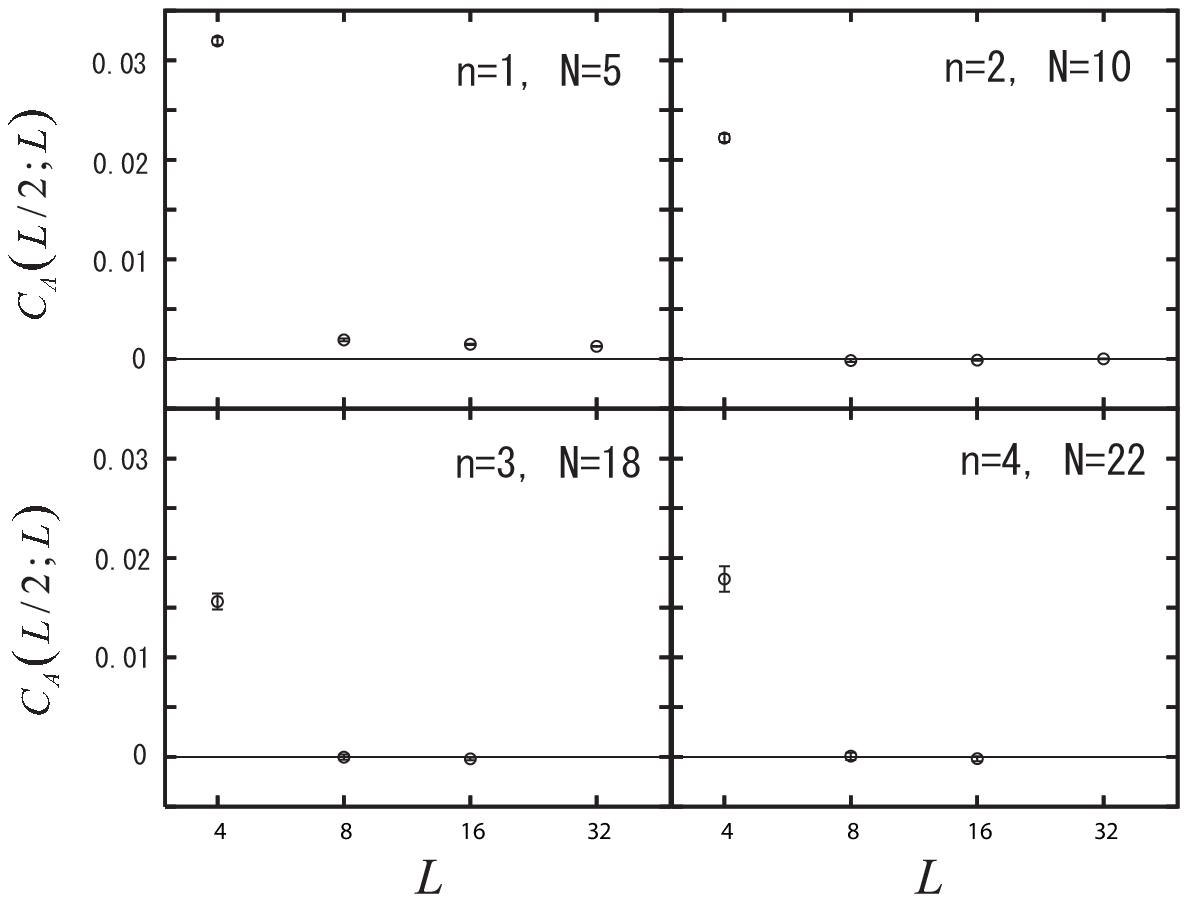}{0.60}
{
The correlation function of the rotational-symmetry-breaking
order-parameter $A$ at the largest distance for $n=1,2,3$ and $4$.
The value of $N$ for each $n$ is close to but larger than the 
ythreshold value.
}

Lastly, we turn our attention to the nature of the non-magnetic
ground states for $n \ge 2$.
As mentioned above, any type of the VBS states in \Fig{VBS} can be
characterized by a non-zero value of $C_A(L/2;L)$ in the $L\to \infty$ limit.
This correlation function is plotted in \Fig{CST_vs_L} for $n=1,2,3$ and $4$.
Characteristic features do not depend on $N$ for each $n$.
We only show the data for a single value of $N$
for the sake of readability of the figure.
(The value of $N$ shown in \Fig{CST_vs_L} is chosen so that 
it is close to but definitely above the estimated threshold value.)
The top left panel for $n=1, N=5$ is shown for comparison.
As is clear from the figure, 
the system possesses the VBS ordering only for $n=1$,
but not for $n=2,3,4$.


To obtain some information as to whether 
the large-$N$ phase at $n \ge 2$ is gapless or not, 
we have also computed the correlation function of 
$P_{\mu}(\vect{R})$ itself,
which is proportional to the energy-energy correlation.
In a clear contrast to the case of $n=1$,
we have seen monotonic decrease to zero for $n=2,3,4$
which is consistent with the absence of the rotational 
symmetry breaking for $n=2,3,4$ mentioned above.
We do not clearly observe exponential decay 
in any of the disordered cases.
The decay in the correlation up to the length scale of $L\sim 32$
is consistent with the power law with the decay exponent between 2 and 3.
However, since the correlation function for large $L$ is very small 
and the relative error is too large to obtain a definite answer to 
the question whether this phase is gapless or not.

To summarize, in addition to the confirmation of the previous
result\cite{HaradaKT2002}, we have found an evidence for an emergent
(exact or approximate) U(1) symmetry of the ground state space of the SU(N) model
with the fundamental representation, and also found no VBS order
in the large $N$ region adjacent to the phase boundary to the N\'eel region.
The latter finding is the first strong evidence for the spin-liquid
(disordered) ground state in the present model that does not have
geometrical frustrations in contrast to the models studied previously
in search for the quantum disordered states.
As for the first finding, it is appropriate to mention the recent works
on the deconfinement critical phenomena (DCP).
The present SU($N$) model is the model that was discussed
in previous works\cite{SenthilETAL2004I,SenthilETAL2004II,SenthilETAL2005}
related to DCP, where the authors argured that some unconventional type of 
the second-order phase transition is possible between 
two phases with apparently unrelated symmetries 
(e.g., the VBS phase and the N\'eel phase).
There also discussed was the emergent U(1) symmetry at 
(or near\cite{SenthilETAL2005}) the critical point.
We suspect that the U(1) structure that we have observed
reflects validity or approximate validity of the DCP scenario
in the present model at zero temperature.

The author would like to thank Cristian Batista and Anders Sandvik for
useful comments.
The computation presented here was performed on SGI Altix 3700/1280 
at Supercomputer Center, Institute of Solid State Physics, 
University of Tokyo.


\end{document}